\documentclass[twocolumn,aps,showpacs,prb,tightenlines,amsmath,amssymb]{revtex4}
\usepackage{graphicx}
\usepackage{amssymb}
\usepackage{dcolumn} 
\usepackage{amsmath}
\usepackage{bm}
\usepackage{colordvi}
\usepackage{mathrsfs}

\makeatletter

\newcommand{\Rmnum}[1]{\expandafter\@slowromancap\romannumeral #1@}
\makeatother

\begin{document} 

\title{Negative differential transmission in graphene} 
\author{B. Y. Sun} 
\author{M. W. Wu}
\thanks{Author to  whom correspondence should be addressed}
\email{mwwu@ustc.edu.cn}
\affiliation{Hefei National Laboratory for Physical Sciences at
  Microscale and Department of Physics,
University of Science and Technology of China, Hefei,
  Anhui, 230026, China}
\date{\today}

\begin{abstract}
By using the Kubo linear response theory with the Keldysh Green function approach,
we investigate the mechanism leading to the negative differential
transmission in system with the equilibrium electron density much smaller than
the photon-excited one. It is shown that the
 negative differential transmission can 
appear at low probe-photon energy (in the order of the
scattering rate) or at high energy (much larger than the scattering rate). 
For the low probe-photon energy case, the negative differential transmission is
found to come from the increase of the intra-band conductivity due to the large
variation of electron distribution after the pumping. As for
the high probe-photon energy case, the negative differential transmission is
shown to tend to appear with the hot-electron temperature
being closer to the equilibrium one and
the chemical potential higher than the equilibrium one but considerably smaller
than half of the probe-photon energy. We also show that this negative
differential transmission can
come from both the inter- and the intra-band components of the
conductivity. Especially, for the inter-band component, its contribution to
the negative differential transmission is shown to come from both 
the Hartree-Fock self-energy and the scattering. Furthermore, the influence of
the Coulomb-hole self-energy is also addressed.   
\end{abstract}


\maketitle

\section{INTRODUCTION}

Graphene is an easily accessible, truly two-dimensional system which is
 attractive from the point of view of both the basic physics and possible
applications. In the past decade, it has attracted immense investigations.\cite{NetoReview,DSarmaReview,Peres,Schwierz,Young,AvourisNanoLett,AbergelAdvPhy,ChoiCRSSMS,Kotov,Castro,Bonaccorso,Avouris} Among
these works, the time-resolved optical pump-probe measurement is widely
used to investigate the dynamics of electrons in graphene.\cite{Dawlaty2008,Wang,Huang,BreusingPRL102,ShangACSNano,GeorgeNanoLett,WinnerlPRL,BreusingPRB83,DSunPRL,LiPRL108,Tielrooij,ChoiApl94,Ruzicka2010,BridaArXiv,NewsonOpt,Hale,Dani2012,Plochocka,Shang2010,Ruzicka2012}
With this method, a pump pulse is used 
to excite electrons from the valence band to the conduction one and then a 
time-delayed probe pulse with the probe-photon energy $\omega$ is
applied to detect the differential transmission (DT). The experimentally
 obtained DT under a  probe-photon energy much higher than twice of
  the Fermi energy is often positive with its
fast relaxation of several hundred femtoseconds followed by a slower picosecond
relaxation.\cite{Dawlaty2008,Wang,Hale,Ruzicka2010} From the DT, the conductivity
$\sigma({\omega})$ is extracted. Moreover, if $\omega$ is also much larger than
  the scattering rate and 
the electrons are considered to be free, the conductivity is given by
($\hbar\equiv 1$)\cite{Dawlaty2008,PeresPRB,Scharf2013}
\begin{equation}
\sigma(\omega)\approx\sigma_{\rm free}(\omega)= -\sigma_0[{f}_{{\bf k_\omega},1}-{f}_{{\bf k_\omega},-1}],
\label{conductivityInter}
\end{equation}
in which $\sigma_0=e^2/4$, ${f}_{{\bf k_\omega},\eta}$ is the electron
  distribution with $\eta=1$ $(-1)$ representing the conduction (valence) band
  and $k_\omega=\omega/(2\hbar v_F)$ stands for the resonant absorption
  state. With this equation and the temporal evolution of 
  $\sigma_{\rm free}(\omega)$, the evolution of the electron distribution difference at the
  resonant absorption state  ${f}_{{\bf
      k_\omega},-1}-{f}_{{\bf k_\omega},1}$ is directly detected. 

In addition to the positive DT, the negative DT, which indicates 
the increase of the
conductivity after the pumping (positive differential conductivity), has also
been observed in different
experiments.\cite{DSunPRL,GeorgeNanoLett,ShangACSNano,BreusingPRB83,BreusingPRL102}
Sun {\it et al.} reported a negative DT in a system of very high equilibrium
Fermi energy, with the probe-photon energy being lower than twice of the Fermi
energy.\cite{DSunPRL}  Soon after that, the negative DT was also
observed by George {\it et al.} in the case where the probe-photon energy is as
low as tens of meV.\cite{GeorgeNanoLett} These two kinds of negative DT can be
well understood theoretically. By considering the inter-band conductivity
of free electrons [Eq.~(\ref{conductivityInter})], the negative DT reported by Sun
{\it et al.}\cite{DSunPRL} is shown to come 
from the weakening of the Pauli blocking due to the heating of electrons by the
pump pulse.\cite{Sun2013} As for the case 
with low probe-photon energy,\cite{GeorgeNanoLett}
 the heating of the electrons can
still contribute to the negative DT.\cite{Romanets2010,WinnerlPRL} Nevertheless, with the scattering
taken into consideration, the intra-band conductivity was found to be also important for
the negative DT.\cite{WinnerlPRL,GeorgeNanoLett} 

In contrast to the previous two
cases with moderate or small probe-photon 
energies, the negative DT has also been observed with the probe-photon energy
being very high (e.g.,
$>1.5$~eV).\cite{ShangACSNano,BreusingPRB83,Plochocka,Shang2010,Ruzicka2012} The
mechanisms lead to this kind of negative DT is still unclear. At    
such high probe-photon energy, the Fermi surface is less possible to be as high
as half of the probe-photon energy and hence the weakening of the Pauli
blocking is less important.  As for the intra-band 
conductivity, its strength is largely
suppressed and hence whether it is strong enough to lead to the negative DT 
is unclear. Moreover, in previous works investigating problems of the
negative DT, the scattering strength used was assumed to be 
constant.\cite{WinnerlPRL} With constant scattering strength,
the inter-band conductivity can not lead to the negative DT. 
Nevertheless, with the
variation of the scattering strength taken into consideration, whether it can
lead to negative DT is still unknown. 
In addition, the variation of the Coulomb self-energy after the 
pumping was also suggested to be the mechanism leading to the increase of the
inter-band conductivity and hence the negative  
DT.\cite{Plochocka,ShangACSNano,BreusingPRB83} However, whether it is
strong enough to lead to a negative DT has  not been fully theoretically
demonstrated and it is even claimed that the negative DT itself may come from
external factors.\cite{Ruzicka2012}

In this work, we calculate the optical conductivity based on the linear response
theory with the Keldysh Green function approach\cite{Altland,Mahan} by taking 
into account the
contribution of the electron-phonon and the electron-impurity scatterings 
as well as the Coulomb
self-energy explicitly. We show that for graphene with low equilibrium
electron density without the Coulomb-hole (CH)
self-energy,\cite{ChowKoch,Choi2001} the negative DT can appear when the
probe-photon energy $\omega$ is low (in the order of the scattering rate) or when 
$\omega$ is high (much larger than the scattering rate). For the low $\omega$
case, the negative DT mainly comes from the large increase of
the intra-band conductivity after the pumping due to the large variation of the
electron distribution. As for the high $\omega$ case, the negative DT is shown
to tend to appear when the hot-electron temperature $T_e$ is close to the equilibrium one
and the chemical potential is higher than the equilibrium chemical potential but considerably 
lower than $\omega/2$. This negative DT is found to come from the variation of
both the intra- and the inter-band components of the conductivity. Especially,
for the inter-band conductivity, we find that both the scattering and the
Hartree-Fock (HF) self-energy can cause the negative DT.\cite{Haug} In addition, the influence of CH
self-energy is also addressed.

This paper is organized as follows. In Sec.~{\Rmnum 2}, we set up the model and
lay out the formalism. In Sec.~{\Rmnum 3} the results obtained numerically  are
presented. We summarize and discuss in Sec.~{\Rmnum 4}.

\section{MODEL AND FORMALISM}

The effective Hamiltonian of graphene
near the $K$ and $K^\prime$ points can be described by\cite{DiVincenzoPRB} 
\begin{equation}
H_0=\sum_{{\bf k}s\mu\eta}\eta\varepsilon_{\bf k}c^\dagger_{{\bf k}s\eta\mu} c_{{\bf k}s\eta\mu}.
\label{EffectHaml}
\end{equation}
Here, $\mu=1$ $(-1)$ stands for $K$ ($K^\prime$)
valley, $\eta=1$ $(-1)$ represents conduction
(valence) band, $s$ denotes spin and $\varepsilon_{\bf k}=v_Fk$ with $v_F$ being the Fermi
velocity. In this work, we only investigate the linearly polarized normal incident
light with the electric field along ${\bf \hat{x}}$ direction. Then, based on the
effective Hamiltonian, the current operator reads (the in-plane photon momentum
$q=0$)\cite{Scharf2013} 
\begin{equation}
{\bf \hat{j}}=\sum_{{\bf k}s\mu\eta\eta^\prime}d^\mu_{\theta_{\bf k}\eta\eta^\prime}c^\dagger_{{\bf k}s\eta\mu} c_{{\bf k}s\eta^\prime\mu},
\end{equation}
with the dipole matrices being
\begin{eqnarray}
d^\mu_{\theta_{\bf k}\eta\eta^\prime}=-|e|v_F[\eta^\prime e^{i\mu\theta_{\bf k}}+\eta
e^{-i\mu\theta_{\bf k}}]/2.
\end{eqnarray}

It has been shown that the hot-electron Fermi distribution is quickly
established after the pumping with the hot electron temperatures in the 
conduction and valence bands being the same while the chemical potentials being
different.\cite{Sun,BreusingPRL102} Based on this result, the electron density matrix without
interactions is set to be $\rho_0=\exp[-\sum_{{\bf
    k}s\mu\eta}\beta_e(\eta\varepsilon_{\bf k}-\mu_\eta)c^\dagger_{{\bf
    k}s\eta\mu} c_{{\bf k}s\eta\mu}]$ with $\mu_\eta$ being the chemical
potential in band $\eta$, $\beta_e=k_BT_e$ and $T_e$ standing for the electron
temperature. Then, from the Kubo formula with the Keldysh Green function approach, the
optical conductivity is expressed as\cite{Altland,Mahan,Lei1992,PeresPRB}
\begin{eqnarray}
\sigma(\omega)&=&\int d{\bf k}\int d\omega_1 g_d\sum_{\eta\eta^\prime}
d^\mu_{\theta_{\bf k}\eta\eta^\prime}d^\mu_{\theta_{\bf k}\eta^\prime\eta}A({\bf k},\eta^\prime,\omega_1+\omega) \nonumber \\
&&\hspace{-1cm}\mbox{}\times A({\bf
  k},\eta,\omega_1)[F(\omega_1,\eta)-F(\omega_1+\omega,\eta^\prime)]/(16\pi^3\omega).
\label{conductivity}
\end{eqnarray}
Here, $g_d=4$ comes from the valley and spin degeneracies and
$F(\omega,\eta)=\{\exp[(\omega-\mu_\eta)/(k_BT_e)]+1\}^{-1}$. In this equation,
the two band indices $\eta$ and $\eta^\prime$ in
$[F(\omega_1,\eta)-F(\omega_1+\omega,\eta^\prime)]$ can be used to distinguish
the contributions of the intra- and inter-band conductivities. The 
intra-band conductivity is defined as terms with $\eta=\eta^\prime$ and the
inter-band one with $\eta\neq\eta^\prime$. With this definition of the
  inter- and intra-band conductivities, their differential conductivities
  are defined as the difference of the corresponding conductivities after
and before  the pumping. The electron spectral function is given by 
\begin{equation}
A({\bf k},\eta,\omega)=-2{\rm Im}\{[\omega-\eta\varepsilon_{\bf
  k}-\Sigma^{\rm R}({\bf k},\eta,\omega)]^{-1}\}.
\end{equation}
Here, $\Sigma^{\rm R}({\bf k},\eta,\omega)$ represents the retarded self-energy
from the electron-electron, electron-phonon and electron-impurity 
interactions. The self-energy from the electron-impurity 
scattering is calculated from\cite{Scharf2013}
\begin{eqnarray}
&&\hspace{-0.5cm}\Sigma_{\rm i}(k)=-\frac{i\pi
  n_i}{2}\int\frac{d^2k^\prime}{(2\pi)^2}\Big(\frac{V_{{\bf k^\prime}-{\bf
    k}}}{\epsilon_{\bf q}}\Big)^2\delta(\varepsilon_{\bf k}-\varepsilon_{\bf k^\prime})\nonumber \\ 
&&\mbox{}\times[1-\cos(\theta_{\bf k}-\theta_{\bf k^\prime})][1+\cos(\theta_{\bf
  k}-\theta_{\bf k^\prime})].
\label{Imselfenergy}
\end{eqnarray}
Here, $\epsilon_{\bf q}$ is the dielectric function calculated as Scharf
{\it et al.}\cite{Scharf2013} and $V_{\bf q}=2\pi e^2 /(\kappa q)$ represents the
two-dimensional bare Coulomb potentials with $\kappa$ standing for the
background dielectric constant.\cite{Scharf2013}

For the self-energy from the
electron-phonon scattering, although the temperatures of electrons and phonons
can be different, the Feynman rules and diagrammatic technique for the Keldysh
Green function approach are still valid.\cite{Lei1992} With this approach,
the retarded self-energy is obtained as 
\begin{eqnarray}
&&\hspace{-0.7cm}\Sigma^{\rm R}_{\rm{ph}}({\bf k},\eta,\omega)
=\sum_{{\bf q},\eta^\prime,\lambda}M_{\eta,\eta^\prime,\lambda}({\bf k}-{\bf q},{\bf q})\nonumber
\\ &&\mbox{}\times
M_{\eta^\prime\eta,\lambda}({\bf k},-{\bf q})\Biggl[\frac{n_{{\bf q}\lambda}+1-F(\epsilon_{|{\bf
      k}-{\bf q}|},\eta^\prime)}{\omega+i0^+-\eta^\prime\epsilon_{|{\bf k}-{\bf
      q}|}-\omega_{\lambda}({\bf q})}\nonumber
\\ &&\mbox{}+\frac{n_{{\bf q}\lambda}+F(\epsilon_{|{\bf k}-{\bf
      q}|},\eta^\prime)}{\omega+i0^+-\eta^\prime\epsilon_{|{\bf k}-{\bf
      q}|}+\omega_{\lambda}({\bf q})}\Biggr].
\label{phononselfenergy}
\end{eqnarray}
Here, $M_{\eta,\eta^\prime,\lambda}({\bf k},-{\bf q})$ are the electron-phonon scattering
matrices for phonons in branch $\lambda$ with their detailed forms given by
Scharf {\it et al.};\cite{Scharf2013} $n_{{\bf
    q}\lambda}=\{\exp[\omega_{q,\lambda}/(k_BT_{\rm ph})]-1\}^{-1}$ is the
distribution for phonons with energy $\omega_{q,\lambda}$ and
temperature $T_{\rm ph}$. In the following, we ignore the effect of
polaronic shifts, i.e., the real part of the electron-phonon
self-energy and compensate such approximation by renormalizing the spectral function
to fulfill $\int d\omega A({\bf k},\eta,\omega)=2\pi$ as done by Scharf {\it et
  al.}.\cite{Scharf2013}

As for the self-energy from the Coulomb
interaction, we take into account the HF self-energy\cite{ChowKoch}
\begin{equation}
\Sigma^{\rm HF}_{{\bf k},\eta}=\sum_{\bf k^\prime}-\eta\cos(\theta_{\bf k}-\theta_{\bf k^\prime})\frac{V_{\bf k^\prime-k}}{2\epsilon_{{\bf k^\prime}-{\bf k}}}(f^e_{\bf k^\prime}+f^h_{\bf k^\prime}).
\end{equation}
Here, $f^e_{\bf k}=F(\varepsilon_{\bf k},1)$ and $f^h_{\bf k}=1-F(-\varepsilon_{\bf
  k},-1)$ are the electron and hole distributions, respectively.
This self-energy comes from the Coulomb exchange self-energy\cite{Li2011}
\begin{eqnarray}
\hat{\Sigma}_{\rm ee}({\bf k},\eta,\omega)&=&-\frac{1}{2}\sum_{\bf k^\prime\eta^\prime}[1+\eta\eta^\prime\cos(\theta_{\bf
  k}-\theta_{\bf k^\prime})] \nonumber\\
&&\hspace{-1.5cm}\mbox{}\times F(\eta^\prime v_F{\bf k^\prime},\eta^\prime){V_{{\bf k^\prime}-{\bf k}}}/\epsilon_{{\bf
    k^\prime}-{\bf k}},
\label{HFselfenergy}
\end{eqnarray}
 with the exchange term of the full valence band subtracted as it is
 already included in the original single-particle energy.\cite{Haug} However,
 it is known in semiconductor optics that there is a Debye shift or the CH self-energy in
 addition to the HF self-energy, which is expressed as\cite{ChowKoch,Ren1999,Choi2001}
\begin{equation}
\Sigma^{\rm CH}_{{\bf k},\eta}=\sum_{\bf q}\eta\cos(\theta_{\bf k}-\theta_{\bf
  k+q})V_{\bf q}(1/\epsilon_{{\bf q}}-1/\epsilon^0_{{\bf q}})/2,
\label{CHselfenergy}
\end{equation}
with $\epsilon^0_{{\bf q}}$ representing the dielectric function at $T_e=0$~K and
$\mu_1=\mu_{-1}=0$~meV. The physics of this CH self-energy is as follows. After the
pumping, the pump-excited electrons change the screening strength. As
a result, the subtracted exchange term of the full valence band also varies
[Eq.~(\ref{CHselfenergy})]. It has been shown in semiconductor optics 
that the CH self-energy leads to a strong renormalization of the
band structure (band-gap shrinkage).\cite{Ren1999,Choi2001} Due to the
similar renormalization of the band structure in graphene, it was speculated 
that this variation of the Coulomb self-energy is responsible for the
negative DT.\cite{BreusingPRB83} In this work, we also calculate the
influence of the Coulomb self-energy to check this speculation.

\section{RESULTS}

From the relation between the transmission and conductivity\cite{Dawlaty2008,Rana2008,Dawlaty2008_2,Falkovsky2008}
\begin{equation}
  T_\omega=|1+N_{\rm 
    lay}\sigma(\omega)\sqrt{\mu_0/\epsilon_0}/(1+n_{\rm ref})|^{-2},
  \label{Trans}
\end{equation}
one finds that the negative DT emerges when the optical conductivity
increases after the pumping. Therefore, instead of calculating the DT, we
investigate the variation of the conductivity before and after the pulse
(differential conductivity)
$\Delta\sigma(T_e,\mu_1,\mu_{-1},\omega)=\sigma(T_e,\mu_1,\mu_{-1},\omega)-\sigma(T_0,\mu^0_1,\mu^0_{-1},\omega)$. Here,
for electrons before the pumping, we choose their temperature $T_0=300$~K. We
  concentrate on the case with the photon-excited electron density much larger
  than the equilibrium one and hence choose the chemical potential $\mu^0_1=\mu^0_{-1}=0$ for convenience. Then, due to the
symmetry between electrons and holes, one always has $\mu_1=-\mu_{-1}$ after the 
establishment of the Fermi distribution. In our calculation, the material
parameters are the same as in Ref.~\onlinecite{Scharf2013} and the impurity
density $n_i=5\times10^{11}$~cm$^{-2}$. The temperature of 
phonons $T_{\rm ph}$ is chosen to be the same as electrons and the substrate is
chosen to be SiO$_2$ unless otherwise specified.

\subsection{Contributions of intra- and inter-band differential conductivities}
In this section, we first give an analytical analysis on the intra- and
inter-band components of the conductivity and then show their contributions to
the total differential conductivity numerically. 
\subsubsection{Analytical analysis on the intra- and inter-band conductivities}
The increase of the total conductivity after pumping may come
  from both the intra- and the inter-band components of the conductivity. For the intra-band 
  component in Eq.~(\ref{conductivity}), it 
  can be simplified by assuming the self-energy $\Sigma^{\rm 
    R}({\bf k},\eta,\omega)={\rm Re}\Sigma_{{\bf k},\eta}+i/(2\tau)$. Further by using
  the condition $\mu_1=\mu_{-1}$ and taking the limit $\omega\rightarrow0$, one has
  \begin{eqnarray}
    \sigma_{\rm intra}(\omega)={e^2}\tau D(T_e,\mu_1,\mu_{-1})/[\pi(1+\omega^2\tau^2)].
    \label{intraSigma}
  \end{eqnarray}
  Here, ${\rm Re}\Sigma_{{\bf k},\eta}$ is the real part of the self-energy
  which comes from the HF and CH self-energies and 
\begin{eqnarray}
  D(T_e,\mu_1,\mu_{-1})&=&2\int^\infty_0 d\varepsilon\varepsilon[-\partial_{\varepsilon}
  F(\varepsilon+{\rm Re}\Sigma_{{\bf k},1},1)]\nonumber\\
  &&\hspace{-1.5
    cm}=2\int^\infty_{\frac{-\mu^\prime}{k_BT_e}}dx{(k_BT_ex+\mu^\prime)}\frac{e^x}{(e^x+1)^2},
\label{Dform}
\end{eqnarray}
with $\mu^\prime=\mu_1-{\rm Re}\Sigma_{{\bf k},1}$. From Eq.~(\ref{intraSigma}),
one finds that the increase of the scattering strength, i.e., the decrease of
$\tau$, leads to the increase (decrease) of the conductivity when the probe-photon energy
$\omega$ is larger (smaller) than $\tau^{-1}$. Moreover, from Eqs.~(\ref{intraSigma}) and 
(\ref{Dform}), it is found that the increase of the electron temperature
  $T_e$ and chemical potential $\mu_1$ as well as the decrease of the
  self-energy ${\rm Re}\Sigma_{{\bf k},1}$  lead to the increase of
  $D(T_e,\mu_1,\mu_{-1})$ and hence $\sigma_{\rm intra}$.

As for the inter-band conductivity, its increase may come from the
  variations of the scattering (the imaginary part of the self-energy), the HF
  self-energy and the CH self-energy (the real part of the self-energy). For the
scattering, it can be shown that the inter-band
conductivity increases with the 
decrease of the scattering strength.\cite{appendix} 
Therefore, the positive differential
conductivity may appear when the scattering strength decreases.
On the other hand, for the HF and CH self-energies, their influences are understood from the simplified
Eq.~(\ref{conductivity}) by neglecting the electron-impurity and electron-phonon
scatterings:
\begin{eqnarray}
&&\hspace{-0.7cm}\sigma(\omega)=\frac{g_d\sigma_0v^2_F}{\omega}\int^\infty_0dk k\delta(\omega-2\varepsilon_{\bf k}+\Sigma^{\rm ee}_{{\bf k},-1}-\Sigma^{\rm ee}_{{\bf k},1})\nonumber \\
&&\hspace{-0.0cm}\mbox{}\times[F(-\varepsilon_{\bf k}+\Sigma^{\rm
  ee}_{{\bf k},-1},-1)-F\big(\omega-\varepsilon_{\bf k}+\Sigma^{\rm ee}_{{\bf
  k},-1},1)]\nonumber \\
&&\hspace{-0.5cm}=[F(-\varepsilon_{{\bf k}_\omega}+\Sigma^{\rm
  ee}_{{\bf k}_\omega,-1},-1)-F(\omega-\varepsilon_{{\bf k}_\omega}+\Sigma^{\rm ee}_{{\bf
  k}_\omega,-1},1)]\nonumber\\
&&\hspace{-0.cm}\times{g_d\sigma_0v^2_Fk_\omega}/\{\omega[2v_F-\partial_k(\Sigma^{\rm ee}_{{\bf
    k}_\omega,-1}-\Sigma^{\rm ee}_{{\bf k}_\omega,1})]\}.
\label{interCH}
\end{eqnarray}
Here, $k_\omega$ satisfies $\omega-2\varepsilon_{{\bf k}_\omega}+\Sigma^{\rm ee}_{{\bf
    k}_\omega,-1}-\Sigma^{\rm ee}_{{\bf k}_\omega,1}=0$ and $\Sigma^{\rm
  ee}_{{\bf k}_\omega,\eta}$ stands for the corresponding self-energy from the
Coulomb interaction (HF and CH self-energies). After the pumping, the
difference of the HF self-energy $\Sigma^{\rm HF}_{{\bf
    k}_\omega,-1}-\Sigma^{\rm HF}_{{\bf k}_\omega,1}$ may increase due to the
increase of the carrier density and that of the CH self-energy
 $\Sigma^{\rm CH}_{{\bf  k}_\omega,-1}-\Sigma^{\rm CH}_{{\bf k}_\omega,1}$ 
can also increase due to the
enhancement of the screening strength. As a result, ${k}_\omega$ may increase, which
 tends to enhance the inter-band conductivity.

\begin{figure}[htb]
  \includegraphics[width=8.7cm]{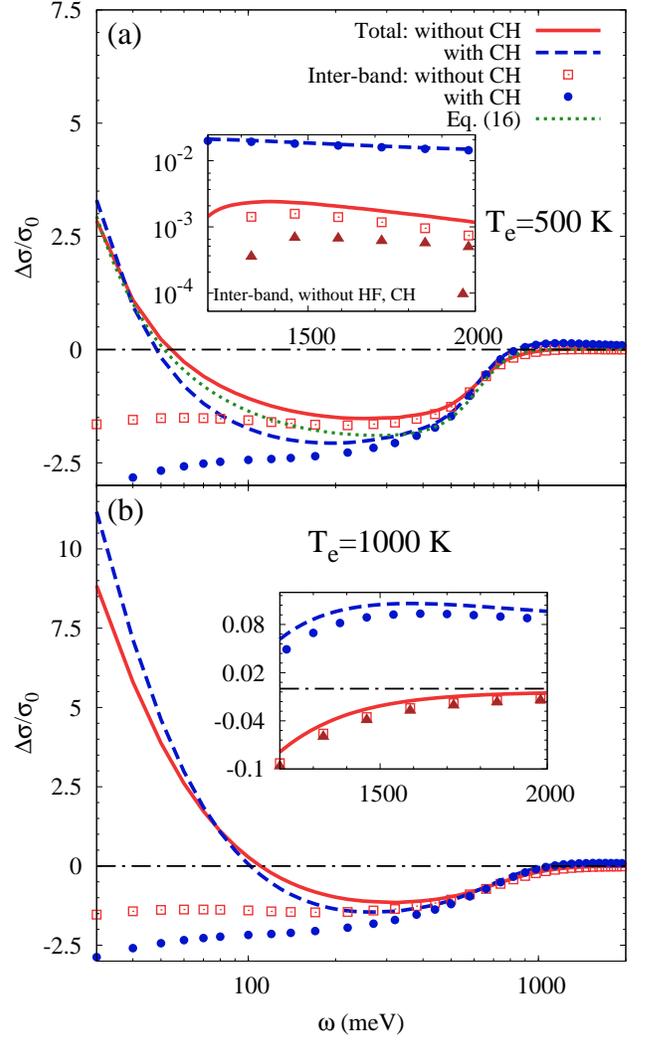}
  \caption{(Color online) Inter-band as well as total differential
    conductivities with and without the CH self-energy as function of the
    probe-photon energy $\omega$. The states after the pumping
are chosen to be (a) $T_e=500$~K and $\mu_1=-\mu_{-1}=300$~meV, and (b)
    $T_e=1000$~K and $\mu_1=-\mu_{-1}=300$~meV. The 
    total conductivity without the CH self-energy in (a) is fitted for
    $\omega\leq50$~meV by Eq.~(\ref{SCFit}) with the fitted results plotted as
    the green dotted curve. The inset zooms the range at high $\omega$ with the
    inter-band differential conductivity without the CH and HF 
    self-energies plotted as brown solid triangles. The black chain line in the
    figure marks $\Delta\sigma/\sigma_0=0$.} 
  \label{figsw1}
\end{figure}

\subsubsection{Numerical investigation on the contributions of intra- and
  inter-band differential conductivities} 
We then numerically calculate the probe-photon energy dependence of the total 
differential conductivity. The case without the CH self-energy is first
investigated. It is noted that the chemical potential $\mu_1$ and
  temperature $T_e$ of hot-electrons vary with the temporal evolution after
  the pumping. Here, we take $T_e=500$~K and $\mu_1=-\mu_{-1}=300$~meV as an
  example to investigate the possible mechanisms discussed in Sec.~IIIA1.
The total differential conductivity as function of probe-photon
energy is plotted in Fig.~\ref{figsw1}(a) (red solid curve). 
To identify the individual contributions of inter- and intra-band components, we
also plot the differential conductivity with only the inter-band component in the
figure (open squares). One finds that, at very low probe-photon energy $\omega$
($<50$~meV), the total differential conductivity is positive and far away from
zero while the inter-band differential conductivity is negative. This indicts
that the large positive differential conductivity at low $\omega$ comes from the
intra-band component.

As analyzed in Sec.~IIIA1, this large positive
intra-band differential conductivity at low probe-photon energy $\omega$ comes
from the variation of $D(T_e,\mu_1,\mu_{-1})$ 
and $\tau$  after the pumping. The variation of
$D(T_e,\mu_1,\mu_{-1})$ can be obtained from 
Eq.~(\ref{Dform}) if the electron distribution (i.e., $T_e$ and $\mu_1$) and the HF self-energy
${\rm Re}\Sigma_{{\bf k},1}$ are known. It is noted that  ${\rm Re}\Sigma_{{\bf k},1}$
for the temperatures and chemical potentials calculated here is always
 less than 3~meV,  much smaller than the chemical potential after the pumping
(300~meV). Hence, the influence of the HF self-energy on
$D(T_e,\mu_1,\mu_{-1})$ can be neglected and one obtains 
$D(T_e,\mu_1,\mu_{-1})=35.8$ and $600$~meV for the cases before and after the
pumping, respectively. To further identify the contribution of $\tau$, we fit
the total conductivity in the region $\omega<50$~meV with   
\begin{equation}
\tilde{\sigma}(\omega)=\sigma_{\rm intra}(\omega)+\sigma_{\rm free}(\omega).
\label{SCFit}
\end{equation}
Here, $\sigma_{\rm free}(\omega)$ is included to take into account the influence of the
inter-band conductivity. $\tau$ is fitted to be 0.07~meV$^{-1}$ 
and 0.17~meV$^{-1}$ for the cases before and after pumping, respectively [the
obtained differential conductivity is plotted in Fig.~\ref{figsw1}(a) as the green
dotted curve]. One finds that the relative change of 
$\tau$ is much smaller than that of $D(T_e,\mu_1,\mu_{-1})$. Hence the
variation of the intra-band conductivity mainly comes from the variation of
the electron distribution, which leads to the huge increase of
$D(T_e,\mu_1,\mu_{-1})$. Moreover, since the influence of the HF self-energy on
$D(T_e,\mu_1,\mu_{-1})$ can be neglected, its influence on the intra-band
component of the differential conductivity is also marginal. 
It is further noted from the figure that, with the increase of the probe-photon
energy $\omega$ from zero, $\Delta\sigma$ decreases fast and becomes negative
around $\omega=55$~meV. This quick decrease of $\Delta\sigma$ can be understood 
from Eq.~(\ref{intraSigma}), which infers that the influence of the intra-band
component of the differential conductivity is significant only when $\omega$ is in
the order of $\tau^{-1}$.

In addition to the large positive differential conductivity shown in the
case of low probe-photon energy $\omega$, another small positive total
differential conductivity also appears when $\omega$ is high (larger than
1200~meV) as shown by the red solid curve in the inset. One also finds from the
inset that the inter-band differential 
conductivity (open squares) is also positive. Moreover, it is smaller than the total one with
their difference being in the same order ($\sim10^{-3}\sigma_0$) as 
the inter-band differential conductivity. This means that the positive total
differential conductivity comes from both the inter- and the intra-band
components and both  contributions are important at such high $\omega$.

For the positive contribution from the intra-band component at such high
$\omega$, it still mainly comes from the large variation of the electron
distribution after the pumping as the case with low probe-photon energy
$\omega$. This is confirmed by calculating the intra-band component in
Eq.~(\ref{conductivity}) with the distribution
[$F(\omega_1,\eta)-F(\omega_1+\omega,\eta)$] fixed with the pumping. We find
 that the obtained intra-band conductivity decreases more
than 75\% for all $\omega$ shown in the inset (not plotted in the figure).

As for the positive contribution from the inter-band component, it may
come from the variations of both the HF self-energy and the scattering
strength  after pumping as discussed in Sec.~IIIA1. Although the
influence of the HF self-energy on the intra-band component is marginal, it may 
have non-negligible influence on the inter-band component. To further identify its
contribution, we also plot the inter-band differential contribution 
without the HF self-energy in the inset (brown solid triangles). 
One finds that the
one with the HF self-energy is about $2\sim3$ times of that without the HF
self-energy. This means that the
variations of both the HF self-energy and the scattering strength have
important contributions to the positive inter-band differential conductivity at
such high probe-photon energy.

We further investigate the influence of the CH self-energy on the differential
conductivity. With the CH self-energy, the
 total differential conductivity and the one with only
inter-band component  are plotted as
the blue dashed curve and the blue solid dots in the 
figure, respectively. One finds that their qualitative behaviors are similar to
those without the CH self-energy, i.e., the large 
positive differential conductivity
from the intra-band component at low probe-photon energy $\omega$ as
well as the emergence of positive differential conductivity at high
$\omega$. Nevertheless, it is seen that the differential conductivity with the
CH self-energy included varies markedly compared with the one without the CH
self-energy. This means that the  CH self-energy has pronounced influence on the differential
conductivity. Moreover, it is also seen that with the CH self-energy, the total and the inter-band
differential conductivities are very close at high $\omega$ as shown in the
inset. This indicates that the large influence of the CH self-energy at high $\omega$ is mainly
on the inter-band component.

To further show the influence of $T_e$, we 
 plot the probe-photon energy $\omega$ dependence of the
  differential conductivities at the electron temperature $T_e=1000$~K and
  chemical potential $\mu_1=300$~meV in Fig.~\ref{figsw1}(b). 
For the case without the CH self-energy, the total
differential conductivity (red solid curve) at high $\omega$ ($>1200$~meV as shown in the
inset) becomes negative, which is different from the case at $T_e=500$~K. To
show the origin of this negative differential conductivity, we  plot the 
 the inter-band one with (red open squares) and without (brown solid
 triangles) the HF self-energy in the inset. It is shown that they are also
 negative and the total differential conductivity is close to the inter-band one
 without the HF self-energy. This indicates that the negative total differential
 conductivity comes from the negative inter-band differential
 conductivity without the CH and HF self-energies under high temperature at high
 $\omega$. To see  the influence of the CH self-energy, we plot  
 in the inset  both the total (blue dashed curve) and the inter-band
(blue solid dots) differential conductivities. One finds both  become positive
for the high $\omega$ case. This again 
shows that the CH self-energy has a large influence on the 
differential conductivity.

\begin{figure}[htb]
  \includegraphics[width=8.7cm,height=11.56cm]{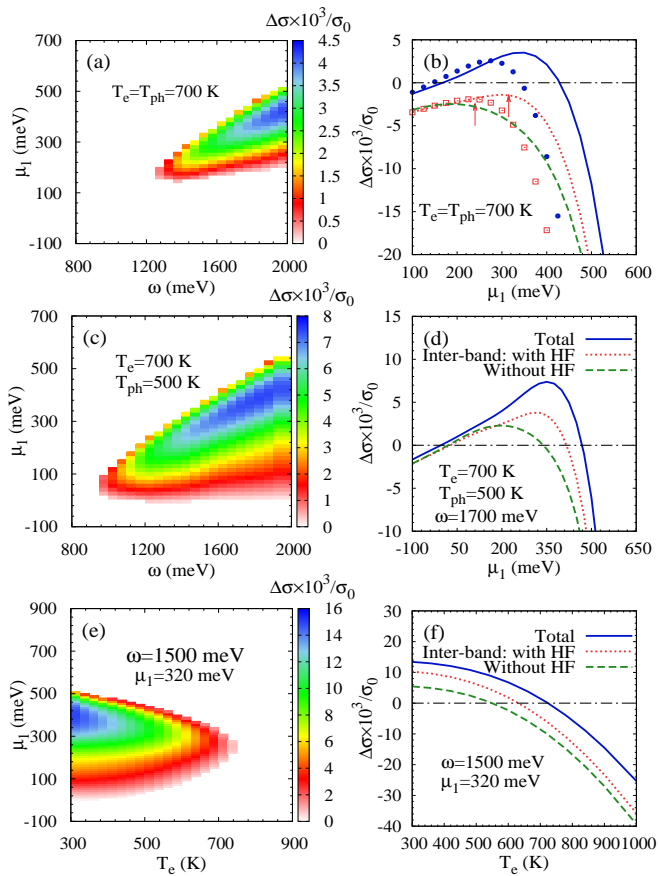}
  \caption{(Color online) (a) Total differential
    conductivity in $\omega$-$\mu_1$ space with $T_{\rm ph}=T_e=700$~K. (b) Chemical potential dependence of the total
    differential conductivity at $\omega=1500$~meV (blue solid dots) and
    1700~meV (blue solid curve) as well as the inter-band one at
    $\omega=1500$~meV (red open squares) and
    1700~meV (red dashed curve). The chemical potential dependence of the
    inter-band differential conductivity without the HF self-energy is also plotted
    as the green dashed curve. Here, $T_e=T_{\rm ph}=700$~K. The red arrows
    mark the peak of the 
    inter-band differential conductivities. (c) Total differential
    conductivity in the $\omega$-$\mu_1$ parameter space with $T_{\rm ph}$
    being 200~K lower than $T_e$. (d) Chemical potential dependence of the total
    differential conductivity at $\omega=1700$~meV as well as that for the
 inter-band one with and without the HF self-energy. Here, $T_e=700$~K and $T_{\rm ph}=500$~K. 
    (e) Total differential conductivities in the $T_e$-$\mu_1$ parameter
    space. Here, $\omega=1500$~meV and $T_{\rm ph}=T_e$. (f) Temperature
    dependence of the total differential conductivity as well as that for the
    inter-band one with and without the HF self-energy. Here, $\mu_1$ is chosen to
    be 320~meV. For clarity, only the positive differential conductivity is
    plotted in (a), (c) and (e). The chain lines in (b), (d) and (f) mark
    $\Delta\sigma/\sigma_0=0$. }  
  \label{figsw2}
\end{figure} 

\subsection{Condition for negative DT}

In the previous section, the mechanisms and their contributions to the positive differential
conductivity are investigated under specific electron temperatures $T_e$ and chemical
potential $\mu_1$. However, the hot-electron temperature and the chemical
potential after the pumping vary with time. Here, we investigate the 
conditions of $T_e$ and $\mu_1$ as well as the probe-photon energy $\omega$ for the
emergence of the positive differential conductivity (i.e., negative DT). Since
the positive differential conductivity in the low $\omega$ case has been
shown to come from the intra-band component of the conductivity, in this
section, we only investigate the case with high $\omega$, where the negative DT was
speculated to come from the scattering  
and the Coulomb self-energy.\cite{Plochocka,ShangACSNano,BreusingPRB83} The cases 
without the CH self-energy are first studied with the total differential
conductivity in the $\omega$-$\mu_1$ parameter space plotted in
Fig.~\ref{figsw2}(a). Here, the electron temperature is chosen to be 700~K,
which is among those estimated in the experiment and the 
phonon temperature is chosen to be the same as that of electrons. We only plot
the region with positive differential conductivity for clarity. It is found
that if the probe-photon energy is larger than 1250~meV, the positive differential conductivity 
emerges with the corresponding range of $\mu_1$ depending on $\omega$. To
further show the contributions of the inter- and intra-band components, we plot
the chemical potential dependence of the total and the inter-band differential
conductivities in Fig.~\ref{figsw2}(b) with $\omega=1500$ and 1700~meV. From the figure, one finds that the inter-band
differential conductivities (open squares and dashed curve) are negative but the 
total conductivity is positive when $160$~meV $<\mu_1<420$~meV for
$\omega=1700$~meV (solid curve) and $140$~meV
$<\mu_1<340$~meV for $\omega=1500$~meV (dots). This indicates that the positive 
differential conductivity comes from the intra-band component. This result is consistent
with previous ones that the intra-band differential conductivity increases
with $T_e$ and $\mu_1$ [Eq.~(\ref{Dform})] while the inter-band one is 
negative if the hot-electron temperature is much larger than the equilibrium
one. Moreover, one also finds that peaks appear in the chemical potential
dependences of both the total and the inter-band differential 
conductivities with the corresponding $\mu_1$ considerablely smaller than
  $\omega/2$. Considering that the intra-band differential conductivity 
increases with $\mu_1$ monotonously, the peak shown in the
total differential conductivity is understood to come from the inter-band one. In addition, it
is also shown in the figure that with the increase of $\omega$, the peak of the 
inter-band differential conductivity appears at higher $\mu_1$ (pointed by the arrows), this
explains that the positive total differential conductivity tends to appear at
higher $\mu_1$ with the increases of $\omega$ as shown in
Fig.~\ref{figsw2}(a). To further show the influence 
of the HF self-energy on the inter-band differential conductivity, we plot the
inter-band differential conductivity at $\omega=1700$~meV without the HF
self-energy (green dashed curve) in Fig.~\ref{figsw2}(b). It is shown that the 
inter-band differential conductivities with and without the HF self-energy are close
at small $\mu_1$ ($<200$~meV). This is understood since the HF self-energy is
small when the excited electron density is small (i.e., $\mu_1$ is small). On the
other hand, when the excited electron density is much larger than the
equilibrium one (i.e., $\mu_1\gg0$), the HF self-energy also becomes large and it leads to an obvious increase
of the inter-band differential conductivity as shown in the figure.

One also finds from Fig.~\ref{figsw2}(a) that the positive differential
  conductivity (negative DT) emerges
  only when the probe-photon energy $\omega$ is high enough. This is supported
 by the experimental data by Shang {\it et al.}.\cite{ShangACSNano}
They showed that the negative DT appears when the 
probe-photon energy $\omega$ is larger than 2000~meV
 but disappears when $\omega$ is smaller than 1835~meV under very strong
 pumping strength. From the numerical results shown in
  Fig.~\ref{figsw2}(a) (where $T_e=700$~K), this property can be shown by
  electrons with the chemical potential $\mu_1$ around 490~meV. One finds that
  the corresponding differential conductivity (DT) changes from negative
  (positive) to positive (negative) when
  $\omega>1900$~meV. It is noted that the electron density
 under these $T_e$ and $\mu_1$ is
  $1.9\times10^{13}$~cm$^{-2}$. This electron density is also reasonable since
  the pumping strength used in the experiment is very strong.

It is noted that in our calculation, all modes of phonons are assumed to be at
the same temperature as electrons. Nevertheless, in genuine situation, the
phonon-phonon scattering is not strong enough to 
buildup such phonon distribution.\cite{Sun,Sun2013,ButscherAPL91}  To further investigate the
influence of the phonon temperature, we also show the results with the phonon
temperature being 200~K lower than the electron temperature $T_e$ in 
Fig.~\ref{figsw2}(c). It is shown that with the decrease of the phonon
temperature, the region of the positive differential conductivity
is enlarged. This is understood since 
the electron-phonon scattering strength decreases with the decrease of the
phonon temperature, which increases the inter-band differential 
conductivity.\cite{appendix} To confirm this,
 we also plot the chemical potential dependence
of the total and the inter-band differential conductivities
with $T_{\rm ph}=500$~K and $\omega=1700$~meV in Fig.~\ref{figsw2}(d). Compared to
the one with $T_{\rm ph}=700$~K [Fig.~\ref{figsw2}(b)], it is found
that the inter-band differential conductivity without the HF self-energy (green
dashed curve) increases markedly and it can even lead to positive differential
conductivity at $45$~meV$<\omega<350$~meV.

We further give a detailed investigation on the total differential conductivity in the
electron-temperature--chemical-potential ($T_e$-$\mu_1$) space at a fixed
probe-photon energy $\omega=1500$~meV. The phonon temperature
is chosen to be the same as the electrons and the results are plotted in
Fig.~\ref{figsw2}(e). One finds that the positive
differential conductivity can appear when $T_e<730$~K. To further identify the
contributions of the inter- and intra-band components, we plot the temperature
dependence of the differential conductivity in Fig.~\ref{figsw2}(f) with
$\mu_1=320$~meV. One finds that the total differential conductivity (solid 
curve) is obviously larger than the inter-band one (dotted
curve). This indicates the important positive contribution of the intra-band
component. Moreover, it is also shown that the difference between 
the inter-band differential conductivities with (red dotted curve) and without
(green dashed curve)  the HF self-energy varies slowly with the temperature. 
This means that in the parameter range investigated here, the influence of
temperature on the HF self-energy is small. In addition, it is also found that 
the inter-band differential conductivity with only scattering 
(i.e., without the HF
self-energy) can also be positive when $T_e$ is close to the equilibrium
  one (300~K). However, it decreases with the increase of the temperature quickly and
becomes negative when $T_e>550$~K.

Now, we can lay out the conditions for the emergence of the negative DT (or
equivalently, positive differential conductivity) in graphene with low
equilibrium electron density. The negative DT can appear both in the low
probe-photon energy $\omega$ (in the order of the scattering rate) case and in
the high $\omega$ case (much larger than the scattering rate). For the low
$\omega$ case, this negative DT mainly comes from the increase of the
intra-band conductivity. This increase is mainly due to the large variation
 of the electron distribution after the pumping and is almost not influenced by the
 HF self-energy in the temperature range investigated here. As for the high
$\omega$ case, although the DT for free electrons [i.e., the one with the
conductivity calculated from 
  Eq.~(\ref{conductivityInter})] can never be negative, the negative DT can
  appear if the scattering is taken into consideration. With the scattering, the
  inter-band conductivity tends to lead to a negative DT when the hot-electron
  temperature $T_e$ is close to the equilibrium one and the chemical potential
  $\mu_1$ is higher than the equilibrium chemical potential but considerably
  smaller than $\omega/2$. Moreover, this region of the negative DT determined by
  the inter-band conductivity can be further expanded by the intra-band one, which always
  tends to decrease the DT if $T_e$  and $\mu_1$ increase after the
  pumping. Furthermore, the HF self-energy can also contribute to the negative
  DT by increasing the inter-band differential conductivity when $\mu_1$
  is much larger than the equilibrium chemical potential (about several hundred
  milli-electron volts larger) in the temperature range usually estimated in the
  experiment (around 1000~K). This further expands the region of the negative DT
  at high $\omega$. We further note that if the phonon temperature is considered to be
smaller than that of electrons, the inter-band differential conductivity
increases, which also expands the region of the negative DT.

\begin{figure}[htb]
  \includegraphics[width=8.7cm]{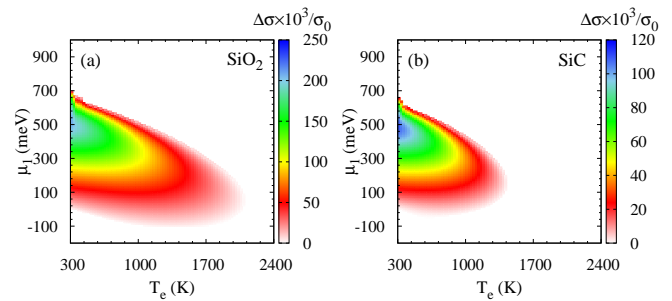}
  \caption{(Color online)  Total differential
    conductivity with the CH self-energy for graphene on SiO$_2$ (a) and  SiC 
(b) in $T_e$-$\mu_1$ parameter space. For clarity, only the 
positive differential conductivity is plotted.}
  \label{figsw3}
\end{figure}

\subsection{Influence of CH self-energy}

Finally, we address the influence of the CH self-energy
on the emergence of the negative DT (positive differential conductivity). The
total differential conductivity with the CH self-energy for graphene on
SiO$_2$ substrate in the electron-temperature--chemical-potential ($T_e$-$\mu_1$) parameter 
spaces is plotted in Fig.~\ref{figsw3}(a). Here, the probe-photon energy
$\omega$ is chosen to be 1500~meV. Comparing with the results without the CH
self-energy [Fig.~\ref{figsw2}(e)], 
one finds that the CH self-energy enhances both the region and the magnitude of
the positive differential conductivity markedly. This  
means that the variation of the CH self-energy after the pumping is the dominant
mechanism leading to the positive differential conductivity at such high
$\omega$ if it is taken into consideration. It is worth noting that, the shape
for the region of the positive differential conductivity with the CH self-energy
is similar to the one without it [Fig.~\ref{figsw2}(e)] (this is also true for
 $\omega$-$\mu_1$ parameter space). This means that the tendency for the
emergence of the total differential conductivity summarized in Sec.~IIIB is
still valid, although the variation of the CH self-energy after the pumping
becomes the dominant mechanism leading to the positive differential
conductivity.

It is noted that since the CH self-energy mainly comes from the variation of the
screening, its influence should be smaller if the graphene
layer is on a substrate with higher background dielectric consistent. To
confirm this, we show the results for graphene on SiC [Fig.~\ref{figsw3}(b)],
whose background dielectric constant is 2.2 times of that for SiO$_2$. One finds
that both the magnitude and the region of the positive differential conductivity decrease
obviously compared with the one on SiO$_2$ substrate. Nevertheless, the positive
differential conductivity (negative DT) region shown on SiC substrate is still 
very large.

Actually, the negative DT region shown on SiO$_2$ and SiC
substrates with the CH self-energy is so large that it disagrees with some of the
existing experiments.\cite{Dawlaty2008,Ruzicka2012,BridaArXiv} 
Dawlaty {\it et al.}\cite{Dawlaty2008} have shown experimentally  for graphene
on SiC substrate that the DT is positive when the optical 
excited electron density is about $10^{12}$~cm$^{-2}$ with the
corresponding $T_e$ estimated to be around 900~K and $\mu_1$ around
zero under the probe-photon energy close to 1500~meV.\cite{Wang} However, the numerically
obtained DT with these parameters is negative as shown in
Fig.~\ref{figsw3}(b). Moreover, Ruzicka {\it et al.}\cite{Ruzicka2012} and Brida
{\it et al.}\cite{BridaArXiv} also showed in graphene on
SiO$_2$ substrate that the DT is positive when the excited electron density
is about $10^{13}$~cm$^{-2}$ with the
corresponding $T_e$ estimated to be around 1000~K and $\mu_1$ around
$350$~meV.\cite{Ruzicka2012} However, the numerically obtained DT with the CH self-energy is
also negative as shown in Fig.~\ref{figsw3}(a). Nevertheless, for the case
without the CH self-energy, the DTs under these parameters are all positive
[Fig.~\ref{figsw2}(e) for the SiO$_2$ substrate and that for the SiC substrate
is not shown]. Further considering the fact that the result obtained without the
CH self-energy is consistent with the experiment with the negative DT
(Sec.~IIIB),\cite{ShangACSNano} one concludes that 
although the calculation of the CH self-energy used here is very successful 
in semiconductors,\cite{ChowKoch,Choi2001} it overestimates the Coulomb
self-energy in the gapless graphene system. This may come from the
 approximations used to obtain Eq.~(\ref{CHselfenergy}).
 For example, the background dielectric constant $\kappa$ is taken to be
constant although it is shown to increase with $q$ [Ref.~\onlinecite{Lischner1302}] and
the vertex correction\cite{Mahan} for the Coulomb screening is
neglected. A better description on the influence of the Coulomb
self-energy is beyond the scope of this investigation.

\section{SUMMARY}
In summary, we have investigated the emergence of the negative DT by calculating
the optical conductivity based on the linear response theory with the Keldysh
Green function approach. In our calculation, the influences of the 
electron-phonon and electron-impurity scatterings as well as the Coulomb
self-energy are explicitly included. We investigate the system with the photo-excited
electron density much larger than the equilibrium one and show that the negative
DT (or equivalently, the positive 
differential conductivity) can appear at low probe-photon energy $\omega$ (in
the order of the scattering rate) or at high $\omega$ (much larger than the
scattering rate).

For the low $\omega$  
case, the negative DT is shown to mainly come from the large increase of
the intra-band conductivity due to the large variation of the electron
distribution after the pumping. As for the high $\omega$ case, it is shown that the 
inter-band conductivity with only scattering can lead to the negative DT when
the hot-electron temperature $T_e$ is close to the equilibrium one and the
chemical potential $\mu_1$ is higher than the equilibrium
chemical potential but considerably lower than $\omega/2$. Moreover, the
intra-band conductivity can further expand this region of the negative DT
because it always tends to decrease DT if $T_e$ and $\mu_1$
increase after the pumping. Furthermore, in the temperature range usually
estimated in the experiments (around 1000~K), the HF self-energy is also shown
to be able to contribute to the negative DT by increasing the inter-band differential 
conductivity considerably when  $\mu_1$ is much higher than the equilibrium
one. This further expands the region of the negative DT at high
$\omega$. Nevertheless, the HF self-energy has little influence on the intra-band
conductivity and hence its effect on the negative DT
at low $\omega$ can be neglected. In addition, we also show that if the phonon
temperature is smaller than that of electrons, the inter-band differential
conductivity is increased which also expands the region of the 
negative DT. Our numerical results are consistent with the $\omega$
dependence of the negative DT obtained experimentally by Shang {\it et
  al.}.\cite{ShangACSNano}

The influence of the CH self-energy is also 
investigated. We find that it markedly expands the region of the negative
DT. However, this expansion is too large which makes the obtained region of the
negative DT become incompatible with the existing
experiments.\cite{Dawlaty2008,Ruzicka2012,BridaArXiv} 
This means that the CH self-energy  calculated with the
well established approach in semiconductors\cite{ChowKoch,Choi2001} 
overestimates the Coulomb self-energy in the
  gapless graphene system. More investigation to take into account 
the high order correlation is needed.

\begin{acknowledgments}
  The authors would like to thank T. S. Lai for valuable discussions. This
  work was supported by the National Basic Research Program of China under Grant
  No. 2012CB922002 and the Strategic Priority Research Program of the Chinese
  Academy of Sciences under Grant No. XDB01000000.
\end{acknowledgments}

\end{document}